\documentclass[a4paper,11pt]{article}
\usepackage[no-natbib-sort]{jinstpub} 

\usepackage{listings}
\usepackage{xcolor}
\definecolor{codegreen}{rgb}{0,0.6,0}
\definecolor{codegray}{rgb}{0.5,0.5,0.5}
\definecolor{codepurple}{rgb}{0.58,0,0.82}
\definecolor{mygreen}{RGB}{28,172,0} 
\definecolor{mylilas}{RGB}{170,55,241}
\definecolor{backcolour}{rgb}{0.95,0.95,0.92}

\graphicspath{{graphics/}}

\usepackage[]{todonotes}

\usepackage{subcaption}

\usepackage{tabularx}

\usepackage{siunitx}

\usepackage[nolist]{acronym}

\usepackage{titlesec}
\titlespacing*{\section}{0pt}{0.5\baselineskip}{0.5\baselineskip}

\usepackage{mytikzpictures/MyTikzPics}

\newsavebox{\imagebox}

\title{\boldmath Design Space Exploration for Particle Detector Read-out Implementations in Matlab and Simulink on the Example of the \acs{SHiP} \acs{SBT}}

\author[1]{F.~Rössing\note{Corresponding Author},}
\author[1]{D.~Arutinov,}
\author[2]{A.~Brignoli,}
\author[3]{H.~Fischer,}
\author[1]{C.~Grewing}
\author[2]{H.~Lacker,}
\author[3]{F.~Lyons,}
\author[1]{A.~Zambanini,}
\author[1,4]{S.~van Waasen}

\affiliation[1]{Central Institute of Engineering, Electronics and Analytics – Electronic Systems (ZEA-2), Forschungszentrum Jülich GmbH, Germany}
\affiliation[2]{Humboldt-Universit\"at zu Berlin, Institut f\"ur Physik, 12489 Berlin, Germany}
\affiliation[3]{Albert-Ludwigs-Universit\"at Freiburg, Physikalisches Institut, 79104 Freiburg, Germany}
\affiliation[4]{Faculty of Engineering, Communication Systems, University of Duisburg-Essen, Duisburg, Germany}
\emailAdd{f.roessing@fz-juelich.de}

\abstract{On a very fundamental level, particle detectors share similar requirements for their read-out chain. This is reflected in the way that typical read-out solutions are developed, where a previous design is modified to fit changes in requirements. One of the two common approaches is the current-based read-out, where the waveform of the sensor output is sampled in order to later extract information in the central processing unit. This approach is used in many detector applications using scintillation based detectors, including \acs{PET}.\\
With this contribution, we will introduce how we use Matlab in order to simulate the read-out electronics of particle detectors. We developed this simulation approach as a base for our ongoing development of software-defined read-out \acp{ASIC} that cover the requirements of a variety of particle detector types. Simulink was chosen as a base for our developments as it allows simulation of mixed-signal systems and comes with built-in toolkits to aid in developments of such systems.\\
With our approach, we want to take a new look at how we approach designing such a read-out, with a focus on digital signal processing closer to the sensor in order or reduce transmission bandwidth, making use of known signal characteristics and modern methods of communications engineering. We are taking into account the time profile of an event, the bandwidth-limiting properties of the sensor and attached electronics, digitization stages and finally the parameterization of approaches for digital processing of the signal. 
We will show how we are applying the design approach to the development of a read-out for the proposed \acs{SHiP} \acs{SBT} detector, which is a scintillation based detector relying on \acs{SiPM}s sensors, using this as an example for our modelling approach and show preliminary results.\\
Given the similarity in principle, the modelling approach could easily be modified for \acs{PET} systems, allowing studies of the read-out chain to possibly gain in overall performance.}

\begin{document}
\subcaptionsetup{skip=5pt}

\begin{acronym}	
\acro{ASIC}[ASIC]{Application Specific Integrated Circuit}
 \acro{SHiP}[SHiP]{Search for Hidden Particles}
 \acro{SBT}[SBT]{Surrounding Background Tagger}
 \acro{ADC}[ADC]{Analog to Digital Converter}
 \acro{TDC}[TDC]{Time to Digital Converter}
 \acro{SPCIE}[SPICE]{Simulation Program with Integrated Circuit Emphasis}
 \acro{ASCI}[ASIC]{Application Specific Integrated Circuit}
 \acro{PET}[PET]{Positron Emission Tomography}
 \acro{DSP}[DSP]{Digital Signal Processing}
 \acro{CERN}[CERN]{European Organization for Nuclear Research}
 \acro{SPS}[SPS]{Super Proton Synchrotron}
 \acro{FPGA}[FPGA]{Field Programable Gate Array}
 \acro{SRAM}[SRAM]{Static-Random Access Memory}
 \acro{WOM}[WOM]{Wavelength-shifting Optical Module}
 \acro{SiPM}[SiPM]{Silicon Photo Multiplier}
 \acrodef{ToT}[ToT]{Time over Threshold}
 \acro{TNR}[TNR]{Trigger-threshold to Noise Ratio}
 \acro{PCB}[PCB]{Printed Circuit Board}
\end{acronym}

\lstdefinestyle{mystyle}{
	backgroundcolor=\color{backcolour}, 
	commentstyle=\color{codegreen},
	keywordstyle=\color{blue},
	numberstyle=\tiny\color{codegray},
	stringstyle=\color{codepurple},
	basicstyle=\ttfamily\scriptsize,
	breakatwhitespace=false, 
	breaklines=true, 
	captionpos=b, 
	keepspaces=true, 
	numbers=left, 
	numbersep=5pt, 
	showspaces=false, 
	showstringspaces=false,
	showtabs=false, 
	tabsize=2,
	aboveskip=\medskipamount
}
\lstset{style=mystyle,language=MATLAB}
\maketitle
\flushbottom
\section{Motivation}
The fundamental characteristics of particle detectors are largely similar. They generate a current pulse, which is the result of the detector`s response to the incident energy, and is influenced by the electrical characteristics of the sensor and read-out electronics. To achieve optimal performance, it is essential to have a comprehensive understanding of the entire system. One effective approach to gain insight into complex systems is to construct a model and test it against real-world observations. This approach is frequently employed in particle physics, utilizing the computational power of modern computers to simulate the behavior of the entire detector and read-out system. In a detector chain, each domain has its own specialized software for such modeling.
The characteristics of an event in the individual channel are typically modeled with tools such as Geant4 \cite{agostinelli2003}, which simulate the particles' behavior in the active material of the detector. The electrical characteristics of the sensor and peripheral electronics and their signal shaping effects are modeled using a derivative of SPICE, an established system for circuit simulations. Finally, the digital post-processing of such signals can be modeled with the tools we already use for digital design, such as Vivado and Cadence.
These tools are highly complex and are used only used by the experts who work on the particular subsystem. In order to derive requirements for our \ac{ASIC} implementations, it is necessary to be able to study the signals and signal processing of a system using a common tool chain. We have chosen Matlab/Simulink to implement a custom approach for studying the mixed-signal processing in particle detectors. In the following, we will provide a brief overview of our approach to modelling and how we are applying it to our developments for a real detector prototype.

\section{Simulation of Particle Detector Read-Out in Matlab and Simulink}
\subsection{Overview}
Matlab and Simulink are standard tools in engineering for the simulation of complex control systems and mixed-signal systems. As such, they are a reliable tool for our developments. The basic concept and central operating principles are outlined in section \ref{sec:framework:architecture} In section \ref{sec:framework:modellingdomains} we will briefly describe how we separated the modeling domains and evaluate their implementations individually. The performance of the simulations will be then discussed in \ref{sec:SimulationPerformance}. In section \ref{sec:SHiP_Simulink} the Simulink model for the \ac{SHiP} \ac{SBT} Detector will be discussed as a case study for both the modelling approach and performance, with the aim of demonstrating the results achieved. These results will be presented in section \ref{sec:simulation_results}. 
\subsection{Architecture}
\label{sec:framework:architecture}
In order to explore parameter spaces for a detector read-out, the ability to perform sweeps of parameters in an efficient manner is essential. This is facilitated by a set of classes and functions that we have implemented in Matlab. The core functionality of repeated parameterized simulations is illustrated in figure \ref{fig:framework:architecture}.
\begin{figure}
	\resizebox{\linewidth}{!}{
		\begin{tikzpicture}
			\pic at(0,0) {framework simple horizontal};
		\end{tikzpicture}
	}
	\caption{Overview of the central architecture our software. It takes care of distribution of parameters, interfacing and parallelization of  simulations. \label{fig:framework:architecture}}
\end{figure}
In a conceptual sense, the software employs a set of parameters to generate a set of events and model configurations. The specific models utilized in event generation and in Simulink must be implemented according to the users requirements, as indicated by the dashed outline of the components in figure \ref{fig:framework:architecture}. 
\subsection{Modelling Domains}
\label{sec:framework:modellingdomains}
Our approach necessitates the investment of modelling effort across multiple fronts. Firstly, the input signal must be modelled in the active part of the detector, which in this context is referred to as event generation. Secondly, the response of the active detector segment to incoming signal quanta must be modelled. This is referred to as sensor modelling. The output of this sensor model is the detector signal current, which is then processed by analog electronics, digitized and further processed by digital signal processing means. These will be referred to as analog and digital processing models.
\paragraph{Event and Sensor Modelling:}
The input to the Simulink model reflects the time characteristics of the detector event. For the simulations performed in the context of this publication we have prepared results from Geant4 simulations. Alternatively custom Monte-Carlo or parameteric approaches can be implemented using the interfaces required by the software.
\paragraph{Simulink Transient Modelling:}
Simulink is a software tool that enables the transient simulation of complex models \cite{themathworksinc.2022}. It facilitates the co-simulation of continuous and discrete time systems, rendering it an optimal choice for simulating both the analog and the digital domains of the signal processing chain for a particle detector. Our models typically comprise a continuous-time component, which models the behavior of analog electronics, and a discrete-time system, which is an implementation of the digital domain logic in Matlab. By deploying our digital logic onto an FPGA using the HDL Coder toolbox, we can test it in a digital hardware environment \cite{mathworks_hdlcoder}. 
The analog domain of our model encompasses both the sensor response to incoming particles and the analog signal processing, including preamplifiers and additional filtering that might be performed for signal shaping. In practice, the user may choose the model complexity and accuracy, as the developed software does not rely on a specific model structure.
As will be discussed later, we primarily utilized the Laplace domain representation (system transfer functions) for expressing and solving the equations of the electrical systems. This approach was selected as it is a straightforward method for modeling continuous-time signals and is well-supported in Simulink. Other potential alternatives include the creation of custom models or the utilization of the Simscape Electrical toolbox. 
In the digital domain, Simulink offers z-transform transfer functions and a plethora of other discrete-time logic blocks via integrated libraries for modeling. This enables the simulation of digital filters, the performance of mathematical operations on the signal, and the application of other kinds of transformations. Moreover, core functionalities of digital systems can be implemented in Simulink, including memory, state machines, and communication interfaces. Simulink supports multi-rate digital systems, allowing the user to take advantage of faster simulations by reducing the sample rate of subsystems through parallelization of data.
\subsection{Simulation Control}
Simulink models can be controlled by Matlab on a run-to-run basis, using objects from the \lstinline{Simulink.SimulationInput} class. These objects allow the parameters of a model to be set and inputs to be provided. The framework developed for this purpose provides functions to automatically create simulation objects and arrange them in a way that allows fast simulations of large datasets. It also utilizes Simulink's built-in precompilation of models and parallelization.
\subsection{Simulation Performance}
\label{sec:SimulationPerformance}
\begin{figure}
	\centering
	\begin{subcaptionblock}[t]{0.49\textwidth}
		\vskip 0pt\centering\captionsetup{width=0.9\linewidth}
		\includegraphics[width=.999\linewidth]{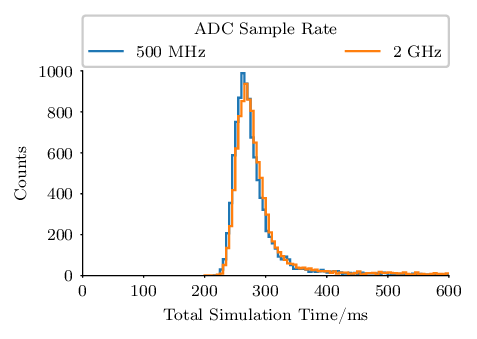}
		\caption{Total time per simulation, including simulation setup and teardown. The mean simulation duration is \SI{299}{\milli\second} for the \SI{500}{\mega\hertz} case and \SI{306}{\milli\second} for the \SI{2}{\giga\hertz}. The 95th percentile is \SI{406}{\milli\second} and \SI{450}{\milli\second} respectively.}
	\end{subcaptionblock}
	\begin{subcaptionblock}[t]{0.49\textwidth}
		\vskip 0pt\centering\captionsetup{width=0.9\linewidth}
		\includegraphics[width=.999\linewidth]{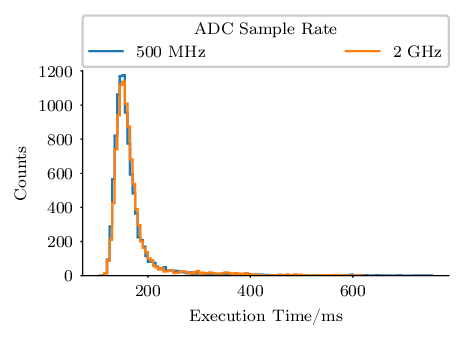}
		\caption{Time taken per simulation, without any pre- or post-processing functions. The mean simulation duration is \SI{166}{\milli\second} for the \SI{500}{\mega\hertz} case and \SI{170}{\milli\second} for the \SI{2}{\giga\hertz}. The 95th percentile is \SI{236}{\milli\second} and \SI{265}{\milli\second} respectively.}
	\end{subcaptionblock}
	\caption{Time performance of simulations using our software, for different ADC sample rates.\label{fig:SimulationPerformance}}
\end{figure}

For the purposes of evaluation, we ran \SI{10000} simulations on an Intel NUC9i7QNX with a 6 Core i7-9750H and 64 GB RAM.
In the context of this evaluation, the model developed for simulation in the \ac{SHiP} \ac{SBT} application context was utilized. The model contains sub-models for the complete analog chain, as well as model of an on-\acs{FPGA} \acs{SRAM} for data buffering, including read and write logic and multiple feature extraction algorithms.
Each simulation covers \SI{1}{\micro\second} of simulated detector signal and contains a single detector event.
The time per simulation is shown in figure \ref{fig:SimulationPerformance}.
The utilization of all cores for parallel simulation results in a reduction of overall run time.
As illustrated by the displayed data, the average simulation takes about \SI{300}{\milli\second}, including approximately \SI{130}{\milli\second} for model setup and tear down functions. The results indicate, that only approximately \SI{75}{\percent} of the total time is utilized for the actual simulation. Consequently, longer multi-event or more complex simulations may be a more attractive option for improving the runtime utilization. Interestingly, increasing the sampling rate by a factor of four has negligible impact on the simulation runtime, indicating, that simulaiton of the digital logic is not the crucial time factor in simulation.
Unfortunately Matlab does not offer other metrics for performance. Using more crude methods of manually checking hardware ressource usage showed a peak CPU utilization of \SI{>93}{\%}, and a peak memory usage of about \SI{18}{\giga\byte}. Hinting, that Simulink comes with a good ressource utilization in terms of CPU time. This points toward further gains in simulation speed per event can only be improved by using more or more performant CPUs for the simulations.
\section{Application to the \acs{SHiP} \acs{SBT} Detector}
In order to demonstrate the accuracy and usefulness of the aforementioned software, we have chosen to apply the approach to an ongoing detector development. The \acl{SHiP} experiment is a particle physics experiment proposed for the ECN3 beam dump facility of the SPS at CERN \cite{bonivento2013,shipcollaboration2022}. The \acl{SBT} sub-detector serves as a background detector for the large vacuum decay vessel, which constitutes the main part of the detector \cite{alt2023}.
\subsection{\ac{SHiP} \ac{SBT} Detector Introduction}
The detector features a long vacuum decay volume, which allows for possible hidden particles to decay with a longer baseline. To track possible intrusions into the decay volume, which could influence the measurements of the downstream detectors, the decay vessel is surrounded by the \ac{SHiP} \ac{SBT}. This detector consists of large steel tanks filled with a liquid scintillator. The individual steel tanks are equipped with \acp{WOM}, absorb the scintillation light and emit it into the lightguide with a shifted wavelength. The light guide is mounted to a \ac{PCB} that hosts a ring of \acp{SiPM}, which convert the collected light into electrical signals.
The ring is made up from 40 individual Hamamtsu s14160-3050HS \cite{hamamatsuphotonics2020} silicon photomultipliers. For readout purposes, five \acp{SiPM} are connected in parallel to form a single read-out channel \cite{shipcollaboration2022,alt2023}. 
The primary objective of the \ac{SBT} detector development is demonstration of the ability to detect interactions. Secondary objectives include quantifying the deposited energy and a position resolution as high as possible. By combining the information from neighboring scintillator tanks, the reconstruction of a particles trajectory can be improved.
The characteristics of photons measured by the \acp{SiPM} are largely dependent on the liquid scintillator, the geometry, and reflectivity of the inner tank walls and photo conversion and collection efficiencies of the \ac{WOM} and \acs{SiPM}.
\subsection{\acs{SHiP} \acs{SBT} Event Model}
\begin{figure}
	\centering
	\includegraphics[width=0.5\textwidth]{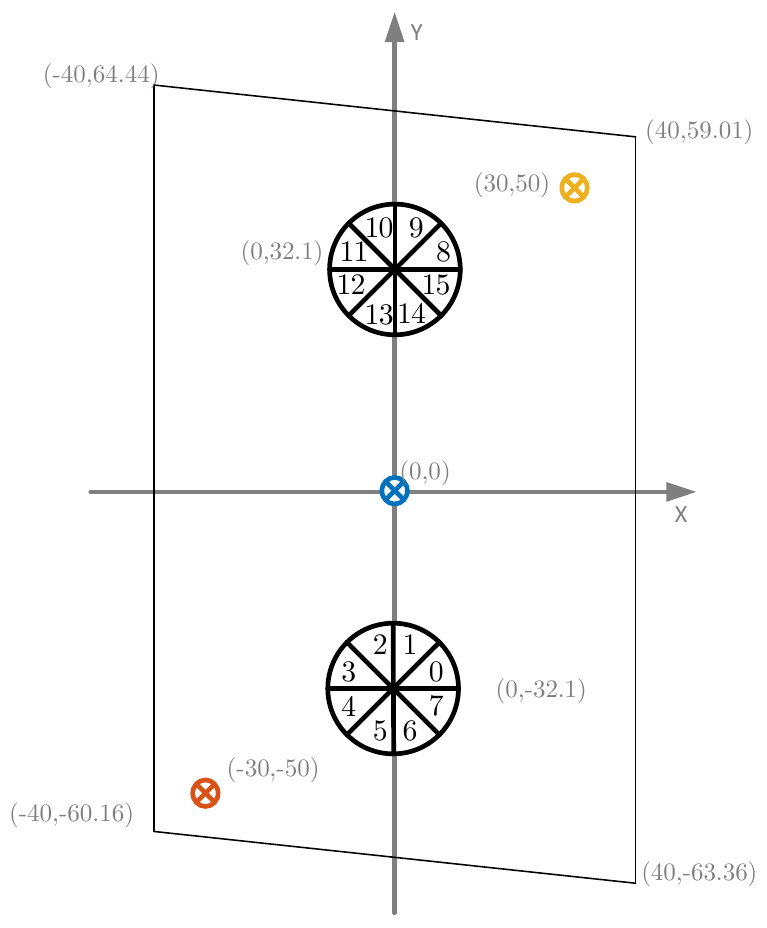}
	\caption{Sketched view of a \ac{SHiP} \ac{SBT} detector cell, with the WOM positions and \acs{SiPM} groups indicated by the cake sliced circles, the numbers in the slices indicate the SiPM groups. The simulated beam positions are indicated by the crossed circles. All coordinates in centimeter, depiction not to scale.} \label{fig:ship_geant_hits}
\end{figure}

\begin{figure}
	\centering
	\begin{subcaptionblock}[t]{0.49\textwidth}
		\vskip 0pt\vskip 0pt\centering\captionsetup{width=0.9\linewidth}
		\includegraphics[width=.999\linewidth]{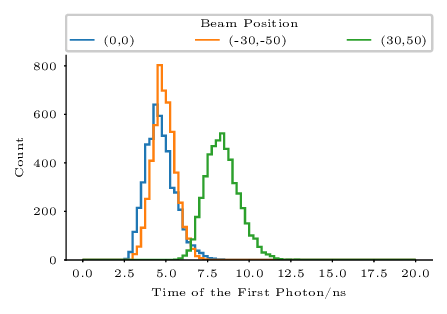}
		\caption{Arrival Time Distribution for the first photon of an event at \acs{SiPM} Group 0. The individual distributions show a standard deviation on the arrival time of \SI{0.7}{\nano\second} with only a small deviation between the hit positions. The mean arrival time shows a strong dependence on the hit position, as can be expected.\label{fig:ship_photonarrival}}
	\end{subcaptionblock}
	\begin{subcaptionblock}[t]{0.49\textwidth}
		\vskip 0pt\vskip 0pt\centering\captionsetup{width=0.9\linewidth}
		\includegraphics[width=.999\linewidth]{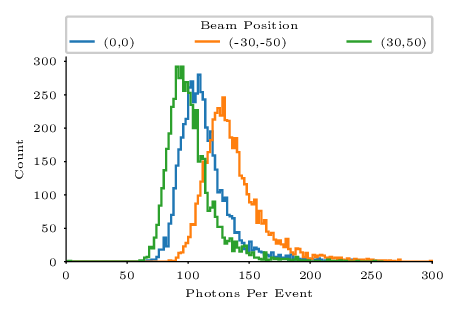}
		\caption{Photon number distributions for photons collected on \acs{SiPM} Group 0 per event. The individual distributions show a a variation if \SI{\sim 20}{percent} w.r.t. their mean. \label{fig:ship_photonnumber}}
	\end{subcaptionblock}
	\caption{Time and photon number resolution for the \ac{SHiP} \ac{SBT} one cell prototype, derived from Geant4 simulations.\label{fig:SHiP_SBT_performance}}
\end{figure}

Given the dependencies on tank geometry and scintillator properties, the simulation of the events is a task that is best left to specialized tools such as Geant4.
For the purpose of this work, we are using the Geant4 model code that was written for the studies of a one-cell prototype detector \cite{lyons}. 
The model accounts for a number of detector characteristics, including scintillation spectra and decay times of the liquid scintillator, wall reflectivity, absorption- and emission spectra of the wavelength shifting layer, the \ac{WOM} transmission, and the wavelength-dependent sensitivity of the \acp{SiPM}.\footnote{We used a tank wall reflectivity of \SI{100}{\percent}, which was determined to be unreasonable in test beam measurements, and the actual reflectivity was measured to be closer to \SI{65}{\percent}.  simulations with the updated Geant4 model and improved electronics will be published at a later stage.}
Geant4 simulations were performed, each containing a single event, with muons of different energies, ranging from \SI{100}{\mega\eV} to \SI{10}{\giga\eV} penetrating the scintillator tank perpendicular to the large face at different positions. Hit positions and tank geometry are displayed in figure \ref{fig:ship_geant_hits}. While we certainly not cover the whole range of possible events expected in operation of the \ac{SHiP} \ac{SBT} sub-detector, the data obtained this way will be suitable as an initial gauge for the usability and reliability of our software. 
From the Geant4 simulation data, we can derive some fundamental limits for the performance of the detector. As illustrated in figure \ref{fig:ship_photonarrival}, the arrival time of the first photon varies by approximately \SI{0.7}{\nano\second} on \acs{SiPM} group 0. We observe a comparable fluctuation on all groups. The combination of the groups yields a significantly improved timing fluctuation of approximately \SI{0.3}{\nano\second}. This sets the boundary for the achievable time resolution accordingly. Given the relatively high amount of dark counts present in SiPMs, triggering of the first photon is unlikely, and we will only be able to achieve a worse time resolution.
With regard to the photon number resolution, shown in figure \ref{fig:ship_photonnumber}, it exhibits a fluctuation of \SI{\pm20}{\percent} with respect to its mean value, thereby limiting the achievable resolution of the photon number estimator.
\subsection{Simulink Model of a single Detector Channel} 
\label{sec:SHiP_Simulink}
The Geant4 output can now be used as the basis for the simulations in our framework. In Matlab, we add dark counts to the simulated event photons, with an adjustable dark count rate; the implementation is similar to \cite{pulko2012}.
In Simulink, we need to emulate the response of a \acs{SiPM} to incoming photons. For this, we employ a modelling approach similar to \cite{grieshaber2022} and \cite{villa2015}, where the modeling of \acp{SiPM} in the context of the \ac{SHiP} \ac{SBT} has already been discussed. 
We use the equivalent circuit, as shown in figure \ref{fig:sipm_equivalent}. We add an idealized model of our front-end electronics to this, so we can account for the shaping effects of the coupling circuitry. The amplifier and coupling components are shown in figure \ref{fig:sipm_and_front-end}.

\begin{figure}
	\centering
	\savebox{\imagebox}{
		\begin{circuitikz}
			\pic at (0,0) {sipm equivalent circuit with labels and boxes};
		\end{circuitikz}
		}%
	\begin{subcaptionblock}[t]{0.49\textwidth}
		\vskip 0pt\vskip 0pt\centering\captionsetup{width=0.9\linewidth}
		\centering\usebox{\imagebox}
		\caption{\acs{SiPM} equivalent circuit model from \cite{grieshaber2022}.\label{fig:sipm_equivalent}}
	\end{subcaptionblock}
	\begin{subcaptionblock}[t]{0.49\textwidth}
		\vskip 0pt\vskip 0pt\centering\captionsetup{width=0.9\linewidth}
		\centering\raisebox{\dimexpr\ht\imagebox-\height}{
			\vspace{1cm}
			\begin{circuitikz}
				\draw[white] (0,0) rectangle ++(5,-7);
				\pic at (0,-1.5) {ship sipm tia};
				\draw[dashed, rounded corners=.3cm] (-.75,-6.5) rectangle ++(1.5,6);
			\end{circuitikz}
		}
		\caption{\acs{SiPM} connected to a trans-impedance front-end.\label{fig:sipm_and_front-end}}		
	\end{subcaptionblock}
	\caption{Equivalent circuitry as used for modelling purposes. For the SHiP SBT single channel model, we use five SiPM structures, highlighted by the dashed box (b), in parallel.}
\end{figure}

\begin{figure}
 \centering
	\begin{subcaptionblock}[t]{0.49\textwidth}
		\vskip 0pt\vskip 0pt\centering\captionsetup{width=0.9\linewidth}
 \includegraphics[width=.999\linewidth]{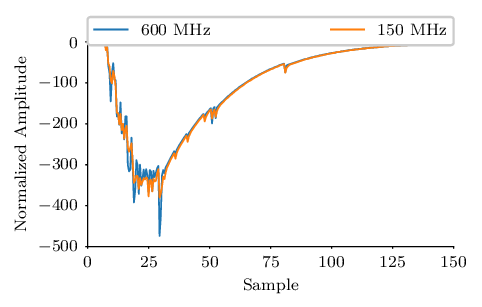}
		\caption{Example waveform from simulations. Higher bandwidths lead to the fast SiPM components becoming prominent and contributing large spikes to the signal, that could potentially reduce resolution. The slower base pulse, is the light curve of the liquid scintillator.\label{fig:WF_v_BW}}
	\end{subcaptionblock}
	\begin{subcaptionblock}[t]{0.49\textwidth}
		\vskip 0pt\vskip 0pt\centering\captionsetup{width=0.9\linewidth}
		\centering
 	\includegraphics[width = 0.999\linewidth]{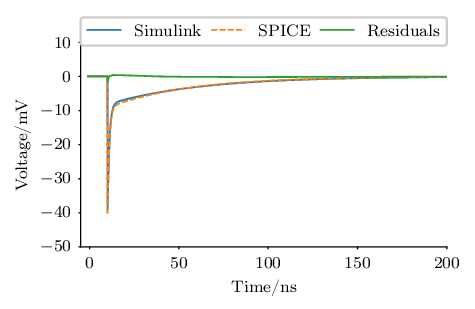}
 	\caption{Comparison of \acs{SiPM} signals captured with the SPICE model from 		\cite{grieshaber2022} and our implementation in Simulink. Residuals are derived from a linear interpolation. The observed residuals are an order of magnitude smaller than the simulated signal.\label{fig:compare_sipm_model}}
	\end{subcaptionblock}
	\caption{Waveforms as obtained from our simulation models. On the left using simulated photon arrivals from a cell of the \ac{SHiP} \ac{SBT}, on the right from a single photo electron.}
\end{figure}

In figure \ref{fig:sipm_equivalent}, the distinction between firing and passive cells of a \acs{SiPM} is made. For a \acs{SiPM} with quenching impedance $R_q$ and $C_q$ and the cell capacity $C_d$, for $N$ total cells and $N_f$ fired cells, the fired cells contribute $R_1=\frac{R_q}{N_f}$, $C_1=C_qN_f$ and $C_2=C_dN_f$. The passive cells $R_3=\frac{R_q}{N-N_f}$, $C_3=C_q\left(N-N_f\right)$ and $C_4=C_d\left(N-N_f\right)$.
In figure \ref{fig:sipm_and_front-end}, we have added the coupling components $R_s$ and $C_c$ that we plan to add in the signal path. We simulate an ideal amplifier, with the frequency limiting pole provided by the feedback structure. The following equations were used to derive our models:
\begin{align}
I_{d}+s C_{2} \left(V_{1}-V_{3}\right)+\frac{V_{1}}{R_{1}}+s C_{1} V_{1}&=0\\
s C_{4} \left(V_{2}-V_{3}\right)+\frac{V_{2}}{R_{3}}+s C_{3} V_{2}&=0\\
I_{d}+s C_{2} \left(V_{1}-V_{3}\right)+s C_{4} \left(V_{2}-V_{3}\right)-C_{5} V_{3}&=s+\frac{V_{3}}{R_{6}}+s C_{c} V_{3}\\
Z_{\mathit{in}}= \frac{1}{s C_{\mathit{c}}};\quad Z_{F}= \frac{1}{\frac{1}{R_{F}}+s C_{F}};\quad\frac{Z_{F} V_{\mathit{in}}}{Z_{\mathit{in}}+Z_{F}}+\frac{Z_{\mathit{in}} V_{\mathit{out}}}{Z_{\mathit{in}}+Z_{F}}&=0;\quad V_{in}=V_3
\end{align}
We can plug the solutions for $V_3$ and $V_{out}$ into a Simulink transfer function block and thus simulate the analog electronics.
We performed a proxy verification by comparing the simulation results from Simulink to the SPICE model from \cite{grieshaber2022} in figure \ref{fig:compare_sipm_model}.

Simulink can be employed to model a multitude of properties associated with an \acs{ADC}, encompassing both basic parameters such as \acs{ADC} bit resolution and intrinsic nonlinearities. The \acs{ADC} output can then be utilized to assess the efficacy of various \ac{DSP} methods in terms of their performance.
For the purposes of this paper, we have tested a few basic approaches.
The most common approach for timing and triggering is to apply a threshold comparison to the signal and record the time of threshold crossing as the signal timestamp. It is typically implemented using a fast comparator and a \acs{TDC}. In this context, we simulate a digital threshold as a baseline algorithm for triggering and timing, as we plan to implement the read-out using only an \acs{ADC}. Parallel to that, we looked at a simple integration of the signal current as an estimator for the total charge created in a detector, and thus for the total deposited energy. Or in the case of the \ac{SHiP} \ac{SBT} the total number of photons collected. As a baseline, we will integrate the signal in the digital domain, by summing up all samples above the threshold, which should give a good estimator for the number of photons. We also looked at sampling the peak of the signal and we recorded the signals \ac{ToT}, which can both serve as a photon number estimator, and also work as a good filter parameter for filtering noise triggers.

\subsection{Simulation Results}
\label{sec:simulation_results}
\begin{figure}
	\centering
	\begin{subcaptionblock}[t]{0.49\textwidth}
		\vskip 0pt\centering\captionsetup{width=0.9\linewidth}
		\includegraphics[width=.999\linewidth]{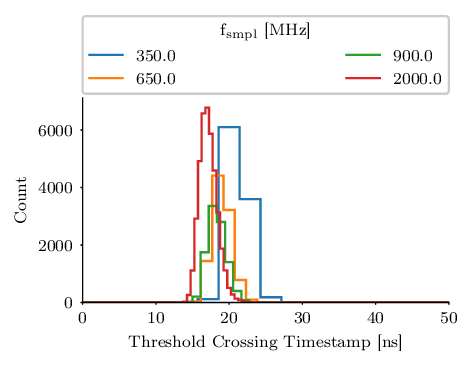}
		\caption{Reported time stamp from the simulink model The binwidth is invers proportional to the sample rate.}
	\end{subcaptionblock}
	\begin{subcaptionblock}[t]{0.49\textwidth}
		\vskip 0pt\centering\captionsetup{width=0.9\linewidth}
		\includegraphics[width=.999\linewidth]{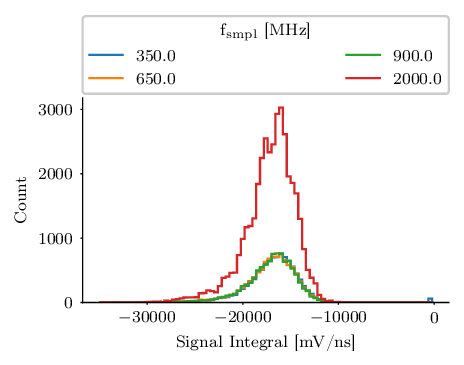}
		\caption{Signal integral as reported by the simulink model.}		
	\end{subcaptionblock}
	\caption{Reported event features from the Simulink model for different ADC sample rates. Some datasets contain more data, as they have been simulated more than others. \label{fig:results_histograms}}
\end{figure}

\begin{figure}
 \centering
 \begin{subcaptionblock}[t]{\textwidth}
 \begin{subcaptionblock}[t]{0.49\textwidth}
 		\vskip 0pt\centering\captionsetup{width=0.9\linewidth}
 		\includegraphics[width=.999\linewidth]{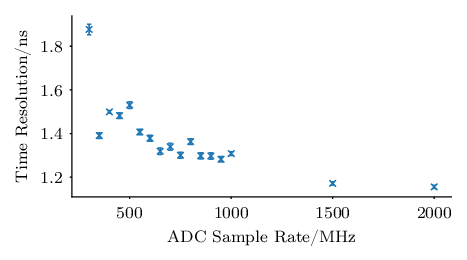}
 		\caption{Time resolution vs. sampling rate.\label{fig:result_TR_SR}}
 	\end{subcaptionblock}
 \begin{subcaptionblock}[t]{0.49\textwidth}
 		\vskip 0pt\centering\captionsetup{width=0.9\linewidth}
 		\includegraphics[width=.999\linewidth]{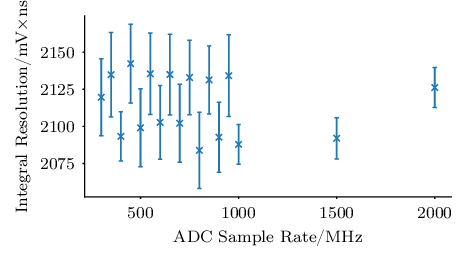}
 		\caption{Integral resolution vs. sampling rate.\label{fig:result_IR_SR}}		
 	\end{subcaptionblock}
 \caption*{Parameterization: Bandwidth$=$ \SI{150}{\mega\hertz}; Trigger threshold $=$ \SI{-30}{\milli\volt}; ADC resolution $=$ \SI{10}{bit}.}
 \end{subcaptionblock}
\begin{subcaptionblock}[t]{\textwidth}
	\begin{subcaptionblock}[t]{0.49\textwidth}
		\vskip 0pt\centering\captionsetup{width=0.9\linewidth}
		\includegraphics[width=.999\linewidth]{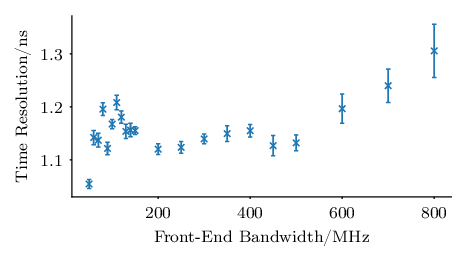}
		\caption{Time resolution vs. bandwidth.\label{fig:result_TR_BW}}
	\end{subcaptionblock}
	\begin{subcaptionblock}[t]{0.49\textwidth}
		\vskip 0pt\centering\captionsetup{width=0.9\linewidth}
		\includegraphics[width=.999\linewidth]{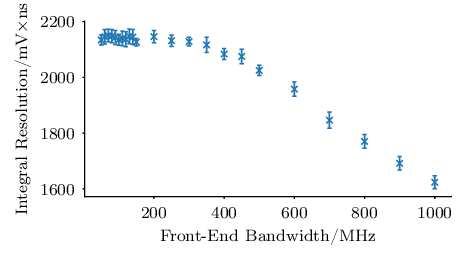}
		\caption{Integral resolution vs. bandwidth.\label{fig:result_IR_BW}}		
	\end{subcaptionblock}
 \caption*{Parameterization: Sample rate$=$ \SI{1}{\giga\hertz}; Trigger threshold $=$ \SI{-30}{\milli\volt}; ADC resolution $=$ \SI{10}{bit}.}
\end{subcaptionblock}
\begin{subcaptionblock}[t]{\textwidth}
 	\begin{subcaptionblock}[t]{0.49\textwidth}
		\vskip 0pt\centering\captionsetup{width=0.9\linewidth}
		\includegraphics[width=.999\linewidth]{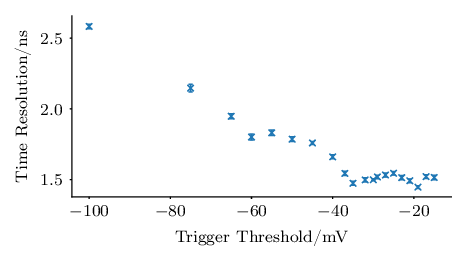}
		\caption{Time resolution vs. trigger threshold.\label{fig:result_TR_thr}}
	\end{subcaptionblock}
	\begin{subcaptionblock}[t]{0.49\textwidth}
		\vskip 0pt\centering\captionsetup{width=0.9\linewidth}
		\includegraphics[width=.999\linewidth]{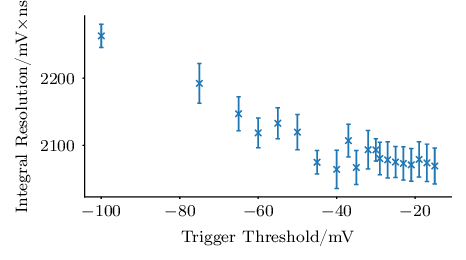}
		\caption{Integral resolution vs. trigger threshold.\label{fig:result_IR_thr}}		
	\end{subcaptionblock}
 \caption*{ Parameterization: Sample rate$=$ \SI{400}{\mega\hertz}; Bandwidth $=$ \SI{150}{\mega\hertz}; ADC resolution $=$ \SI{10}{bit}.}
 \end{subcaptionblock}
 \caption{Influence of different parameters on the estimated timestamp and integral, regarding only cases where no noise was present in the system. The error bars denote the variation introduced by the different parameterizations that were simulated. Only noiseless cases with no dark counts were regarded for the plots.\label{fig:results1}}
\end{figure}

\begin{figure}
 \centering
 \begin{subcaptionblock}[t]{0.49\textwidth}
		\vskip 0pt\centering\captionsetup{width=0.9\linewidth}
		\includegraphics[width=.999\linewidth]{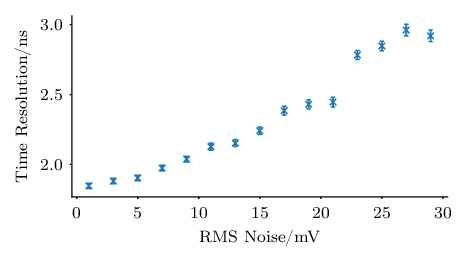}
		\caption{Time resolution vs. RMS Noise.\label{fig:result_TR_noise}}
	\end{subcaptionblock}
	\begin{subcaptionblock}[t]{0.49\textwidth}
		\vskip 0pt\centering\captionsetup{width=0.9\linewidth}
		\includegraphics[width=.999\linewidth]{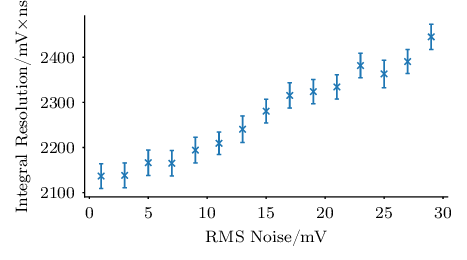}
		\caption{Integral resolution vs. RMS Noise.\label{fig:result_IR_noise}}		
	\end{subcaptionblock}
 \caption{Influence of the RMS noise on time and integral resolution, normalized by the applied trigger threshold. }
 \label{fig:results2}
\end{figure}

In order to narrow down the parameter space for a possible ASIC implementation, we looked at multiple parameter influences. For each simulation run, we simulate a single model parameterization with \si{10000} different events, taken from our Geant4 datasets. We have kept a trans-impedance of \SI{500}{\ohm} and an \acs{ADC} dynamic range of \qtyrange{-0.3}{0.3}{\volt}\footnote{We chose these parameters because they matched the ADC we plan to use for verification.}, and fixed the \acs{SiPM} parameters according to \cite{grieshaber2022}. 
As a first exploration of the parameter space a variety of parameter sweeps were performed, by using a default model configuration and sweeping a single model parameter against that parameterization. From these we obtain histograms of our observables for different parameter values. Examples of those histogramms are shown in figure \ref{fig:results_histograms}. For every parameterization an appropriate distribution is fit to the histogram, using bootstraping \cite{Solr-126059}. In figures \ref{fig:results1} and \ref{fig:results2} we report the fitted standard deviation of the fit, with the parameter uncertainty derived from the bootstrapping results.

The figures \ref{fig:result_TR_SR} and \ref{fig:result_TR_BW} show how the distribution width of the first timestamp is affected by the sample rate and the front-end bandwidth. The influence of the ADC sample rate seems to be small, compared to effort. From a reconstructed timing resolution of \SI{1.4}{\nano\second} to \SI{\sim 1.2}{\nano\second} would require a fourfold increase in sample rate, leading to a highly increased complexity in implementation.

Contrary to usual cases, we observe a degradation in time resolution with increasing bandwidth, as the influence of the high frequency components of the SiPM signals become more dominant than the slower underlying scintillator light curve. Especially for low bandwidths we observe a high, unecpected fluctuation in between  bandwidths. We attribute this behavior to a not ideal fit model, for the distribution of timestamps. A gaussian distribution was used to fit the histogram, while espacially for high bandwidth we observed that the histogram showed two prominent extrema in close proximity, indicating that there are two different effects that can contribute to the signal rise time. This indicates that also for low bandwidths a non-symetric distribution has to be chosen. This systematic effect will be investigated further, later on.
Parallel to that, figures \ref{fig:result_IR_SR} and \ref{fig:result_IR_BW}, show the behavior of the integral resolution. The integral resolution overall seems to be less sensitive to changes in the sample rate and shows an improvement towards higher bandwidths. 
We have observed a negligible influence of the ADC bit resolution on both timestamp and integral resolution.

Figures \ref{fig:result_TR_thr} and \ref{fig:result_IR_thr} show the dependence on the trigger threshold. A clear dependence on the trigger threshold is evident for both observables. In the timestamp domain, we see a jump below \SI{25}{\milli\volt}, which stems from the trigger now being sensitive to individual photons, with the single photon signal peaking at about \SI{20}{\milli\volt}, as shown in figure \ref{fig:compare_sipm_model}. We also observe two steps at approximately \SI{-30}{\milli\volt} and \SI{-60}{\milli\volt} that we can not fully explain yet.
From the plots, we can conclude, that the trigger threshold should be chosen as low as possible. However, choosing a threshold too close to the single photon peak would increase the amount of triggers, hence a compromise has to be made. More investigations, including the effect of a high dark-count rate, have to be performed here.

\begin{figure}
 \centering
 \begin{subcaptionblock}[t]{0.49\textwidth}
		\vskip 0pt\centering\captionsetup{width=0.9\linewidth}
		\includegraphics[width=.999\linewidth]{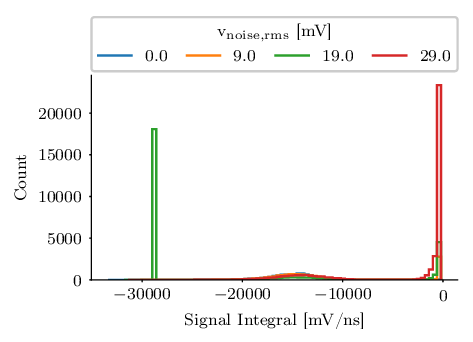}
		\caption{Unfiltered, reported signal integral for different rms noise values.\label{fig:result_noise_histogram}}
	\end{subcaptionblock}
	\begin{subcaptionblock}[t]{0.49\textwidth}
		\vskip 0pt\centering\captionsetup{width=0.9\linewidth}
		\includegraphics[width=.999\linewidth]{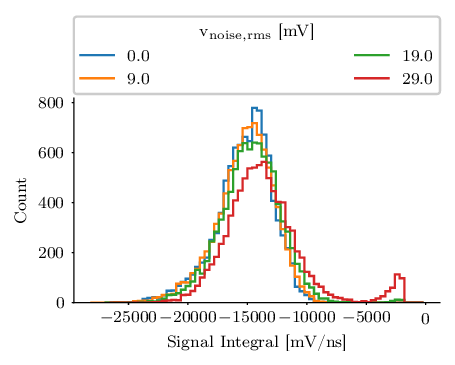}
		\caption{The same data after application of the filters described in the text.\label{fig:result_noise_histogram_filtered}}		
	\end{subcaptionblock}
 \caption{Reported signal integrals, filtered and unfitlered. High noise causes a lot of spurious misstriggers. Through proper datafiltering these can be largely omitted.\label{fig:results_histograms2}}
 
\end{figure}

We also looked at the effect of RMS noise. In order to obtain clean data, the reported data from the model had to be filtered. As is shown in figure \ref{fig:results_histograms2}, a series of cuts were applied. Events where the reported peak exceeded the maximum of the ADC digitization range were excluded, as were events with a \ac{ToT} of less than \SI{25}{\nano\second} or an integral of more than \SI{-2}{\volt\times\nano\second}.
The results of fits to the observables distributions are shown in figures \ref{fig:result_TR_noise} and \ref{fig:result_IR_noise}. 
We used band-limited white noise by adding it to the front-end output. Both time and integral resolution show the expected degradation of resolution with increasing noise, as shown in figures \ref{fig:result_TR_noise} and \ref{fig:result_IR_noise}. 
\section{Comparison to Existing Approaches}
There are many different approaches to simulate parts of a particle detector system. Most prominent the Monte-Carlo simulators for the phyiscal processes, such as Geant4. Additionally there are models for sensor effects, such as \cite{pulko2012} or equivalent circuit models such as reported in \cite{rivetti2015} and \cite{villa2015}. There are also mathematical signal models such as used in \cite{jokhovets2019}. While we use all those individual models, we are not aware of a system that, like our own implementation, stitches together these models to analyse the performance of a complete read-out chain by performing repeated transient simulations.
Typical approaches use the outcomes of individual investigations of the subcomponents to smear distributions such as those shown in figures \ref{fig:results_histograms} and \ref{fig:results_histograms2} to achieve an estimation of achievable performance. 
Our approach reaches beyond this by modelling more of the electronics characteristics and digital logic, giving a more indepth picture.
Unfortunately there was no room for comparative studies to be performed in parallel to our work.
\section{Conclusion and Outlook}
With this work we wanted to present the basic concepts of a software framework for simulating the effects of particle detector read-out electronics on the observables that are of interest for the physics analysis, together with application to the \ac{SHiP} \ac{SBT} detector. 
As demonstrated above the approach is able to perform numerous simulations in a relatively short time frame. By combining models from the different domains of a particle detector, ranging from sensor responses to analog electronics for shaping to digital signal processing, a read-out system can be studied across parameters in all those domains. While the models have to be implemented for each individual detector implemention, the combination of the models in a single framework allows for fine-tuning of the entire single channel system. 

Using the SHiP SBT as an example, we have shown that we can produce meaningful data, that can be used to make predictions for possible hardware implementations.
We are already working on an updated model that has improved analog electronics models and conduct simulations using an adjusted Geant4 model.

To further exploit these capabilities, implementation of optimization algorithms around the above described funcitonalities would be a logical next step. Given our detector agnostic functions, we can easily develop models for a \acs{PET} detector and use the same software to perform design space explorations for a \acs{PET} read-out system.

In the future, we will use our software for the development of possible techniques and algorithms that could help with the improvement of the performance of the read-out for the \ac{SHiP} \ac{SBT} detector. Especially the time resolution could potentially be improved using a suitable interpolation approach, similar to \cite{jokhovets2019}. We plan to verify our simulations with live measurements on a \ac{SHiP} \ac{SBT} detector cell in 2024.

\clearpage
\acknowledgments{We want to thank our collaborators from within the SHiP Collaboration for the support of this work. Espacially we thank Andrii Kotenko and Vladyslav Orlov for important contributions to the Geant4 Model of the SBT detector cell.}

\bibliography{bib}
\end{document}